# Fractal pattern formation at elastic-plastic transition in heterogeneous materials

**J. Li** and **M. Ostoja-Starzewski**

Department of Mechanical Science and Engineering,

Institute for Condensed Matter Theory, and Beckman Institute

University of Illinois at Urbana-Champaign

Urbana, IL 61801

E-mails: junli3@uiuc.edu, martinos@uiuc.edu

**Abstract**

Fractal patterns are observed in computational mechanics of elastic-plastic transitions in two models of linear elastic/perfectly-plastic random heterogeneous materials: (1) a composite made of locally isotropic grains with weak random fluctuations in elastic moduli and/or yield limits; and (2) a polycrystal made of randomly oriented anisotropic grains. In each case, the spatial assignment of material randomness is a non-fractal, strict-white-noise field on a $256 \times 256$ square lattice of homogeneous, square-shaped grains; the flow rule in each grain follows associated plasticity. These lattices are subjected to simple shear loading increasing through either one of three macroscopically uniform boundary conditions (kinematic, mixed-orthogonal or traction), admitted by the Hill-Mandel condition. Upon following the evolution of a set of grains that become plastic, we find that it has a fractal dimension increasing from 0 towards 2 as the material transitions from elastic to perfectly-plastic. While the grains possess sharp elastic-plastic stress-strain curves, the overall stress-strain responses are smooth and asymptote toward perfectly-plastic flows; these responses and the fractal dimension-strain curves are almost identical for three different loadings. The randomness in elastic moduli alone is sufficient to generate fractal patterns at the transition, but has a weaker effect than the randomness in yield limits. In the model with isotropic grains, as the random fluctuations vanish (i.e. the composite becomes a homogeneous body), a sharp elastic-plastic transition is recovered.

# 1. Introduction

It is well known that many materials display fractal characteristics, e.g. (Mandelbrot, 1982, Feder, 2007). Indeed, fractals have been used in the characterization as well as morphogenesis models of spatial patterns. Numerous such phenomena, both in natural and artificial materials, include phase transitions and accretion (e.g. Sornette, 2004), fracture surfaces (Sahimi & Goddard, 1986; Sahimi, 2003; Saouma & Barton, 1994; Shaniavski & Artamonov, 2004), and dislocation patterns (Zaiser *et al.*, 1999). Of course, this is but a short list of such studies, which were extensively conducted in the eighties and nineties.

It appears that very little work was done on fractals in elasto-plasticity, except for (Ostoja-Starzewski, 1990) on plastic ridges in ice fields and (Poliakov *et al.*, 1994; Poliakov & Herrmann, 1994) on shear bands in rocks of Mohr-Coulomb type, Thus, the present paper's focus is on elastic-plastic transitions in planar random materials made of linear elastic/perfectly-plastic phases of metal type. We ask three questions: (a) Does the elastic-plastic transition occur as a fractal, plane-filling process of plastic zones under increasing, macroscopically uniform applied loading? (b) What are the differences between a composite made of locally isotropic grains and a polycrystalline-type aggregate made of anisotropic grains? (c) To what extent is the fractal character of plastic zones robust under changes of the model such as the change of perturbations in material properties?

In this paper we consider elastic/perfectly-plastic transitions in random media in two, two-dimensional microstructural models: (1) a linear elastic-perfectly plastic material with isotropic grains having random yield limits and/or elastic moduli, and (2) a ploycrystal with anisotropic grains following Hill's yield criterion and having random orientations. In both cases, the microstructures are non-fractal random fields, the reason for that assumption being that the evolution of plastic zones would obviously (or very likely) be fractal should the material properties be fractally distributed at the outset. By setting up three types of monotonic loadings consistent with the Hill-Mandel condition, stress-strain responses are numerically obtained and directly related to fractal dimensions of evolving sets of plastic grains. As we observe that the elastic-plastic transition occurs through a fractal for both models, plane-filling pattern of plastic grains, we study the robustness of this result for several related cases.

## 2. Model formulation

By a random heterogeneous material we understand a set $\mathbf{B} = \{B(\omega); \omega \in \Omega\}$ of deterministic media $B(\omega)$, where $\omega$ indicates a realization and $\Omega$ is an underlying sample space (Ostoja-Starzewski, 2008). The material parameters of any microstructure, such as the elasticity tensor or the yield tensor, jointly form a random field $\mathbf{\Theta}$ which is required to be mean-ergodic on (very) large scales, that is

$$\overline{\mathbf{G}(\omega)} \equiv \lim_{L\to\infty} \frac{1}{V} \int_V \mathbf{G}(\omega, \mathbf{x}) dV = \int_\Omega \mathbf{G}(\omega, \mathbf{x}) dP(\omega) \equiv \langle \mathbf{G}(\mathbf{x}) \rangle \tag{1}$$

Here the overbar indicates the volume average and $\langle\ \rangle$ means the ensemble average.

Key issues in mechanics of random materials revolve around effective responses, scales on which they are attained, and types of loading involved. For linear elastic heterogeneous materials, a necessary and sufficient condition of the equivalence between energetically ($\overline{\boldsymbol{\sigma}:\boldsymbol{\varepsilon}}$) and mechanically ($\overline{\boldsymbol{\sigma}}:\overline{\boldsymbol{\varepsilon}}$) defined effective responses leads t the well-known Hill (-Mandel) condition (Hill, 1963) $\overline{\boldsymbol{\sigma}:\boldsymbol{\varepsilon}} = \overline{\boldsymbol{\sigma}}:\overline{\boldsymbol{\varepsilon}}$. As is well known, this equation suggests three types of uniform boundary conditions (BCs):

(1) kinematic (displacement) BC (with applied constant strain $\boldsymbol{\varepsilon}^0$):

$$\mathbf{u} = \boldsymbol{\varepsilon}^0 \cdot \mathbf{x}, \quad \forall \mathbf{x} \in \partial B_\delta; \tag{2}$$

(2) traction (static) BC (with applied constant stress $\boldsymbol{\sigma}^0$):

$$\mathbf{t} = \boldsymbol{\sigma}^0 \cdot \mathbf{n}, \quad \forall \mathbf{x} \in \partial B_\delta; \tag{3}$$

(3) mixed-orthogonal (or displacement-traction) BC:

$$(\mathbf{t} - \boldsymbol{\sigma}^0 \cdot \mathbf{n}) \cdot (\mathbf{u} - \boldsymbol{\varepsilon}^0 \cdot \mathbf{x}) = 0, \quad \forall \mathbf{x} \in \partial B_\delta. \tag{4}$$

Note here that an unambiguous way of writing (4) involves orthogonal projections (Podio-Guidugli, 2000)

$$\left(\mathbf{I} - \mathbf{n} \otimes \mathbf{n}\right) \cdot \left(d\mathbf{u} - d\mathbf{u}^0\right) = 0, \quad \left(\boldsymbol{\sigma}(\mathbf{u}) \cdot \mathbf{n} - \mathbf{t}^0\right) \cdot \mathbf{n} = 0. \tag{5}$$

The above boundary conditions may be generalized to elastic-plastic materials in an incremental setting. Strictly speaking, the traction BC (3) is ill-posed for a perfectly-plastic material, but all the materials in our study are heterogeneous, so that the overall stress-strain responses will effectively be hardening-type for monotonic loadings. We return to this issue in Section 3.

The microstructures in our study are linear elastic/perfectly-plastic materials with an associated flow rule. Specifically, the constitutive response of any grain [i.e. a piecewise-constant region in a deterministic microstructure $B(\omega)$] is described by

$$\begin{aligned} d\boldsymbol{\varepsilon} &= \mathbf{D}^{-1} d\boldsymbol{\sigma} + \dot{\lambda} \frac{\partial f_p}{\partial \boldsymbol{\sigma}} \quad \text{when} \quad f_p = 0 \text{ and } df = 0, \\ d\boldsymbol{\varepsilon} &= \mathbf{D}^{-1} d\boldsymbol{\sigma} \quad \text{when} \quad f_p < 0, \text{ or } f_p = 0 \text{ and } df < 0, \end{aligned} \tag{6}$$

where $\mathbf{D}$ is the elasticity tensor, and $f_p$ is the yield function. For anisotropic materials with quadratic yielding, $f_p$ is taken in von Mises' form

$$f_p = \Pi_{ijkl} \sigma_{ij} \sigma_{kl} - 1. \tag{7}$$

Here $\mathbf{\Pi}_{ijkl}$ represents a positive defined fourth-order tensor of plastic moduli with the following symmetries

$$\Pi_{ijkl} = \Pi_{jikl} = \Pi_{ijlk} = \Pi_{klij} \tag{8}$$

It follows that $\mathbf{\Pi}_{ijkl}$ has only 21 independent components instead of 81 components in the most general case. Two special forms of (7) will be employed:

(i) Huber-von Mises-Hencky (isotropic) yield criterion:

$$f_p = \frac{1}{6}\left[ \left(\sigma_{11} - \sigma_{22}\right)^2 + \left(\sigma_{11} - \sigma_{33}\right)^2 + \left(\sigma_{22} - \sigma_{33}\right)^2 \right] + \sigma_{12}^2 + \sigma_{13}^2 + \sigma_{23}^2 - \frac{\sigma_0^2}{3}. \tag{9}$$

(ii) Hill (orthotropic) yield criterion:

$$f_p = F\left(\sigma_{11} - \sigma_{22}\right)^2 + G\left(\sigma_{11} - \sigma_{33}\right)^2 + H\left(\sigma_{22} - \sigma_{33}\right)^2 + 2L\sigma_{12}^2 + 2M\sigma_{13}^2 + 2N\sigma_{23}^2 - 1. \tag{10}$$

# 3. Computational simulations of elastic-plastic transitions

We consider two special models of such random heterogeneous materials. One consists of isotropic grains and the other is an aggregate of anisotropic grains (crystals). In both cases, the grains are homogeneous, linear elastic/perfectly-plastic with the flow rule following associated plasticity. The Huber-von Mises-Hencky yield criterion applies to the isotropic case, while for the crystals we employ Hill's quadratic orthotropic yielding.

**Model 1.** Isotropic grains with random perturbations in the elastic modulus and/or the yield limit. It follows that the random field of material properties is simplified to $\mathbf{\Theta} = \{E, \sigma_0\}$, in which $E$ is the elastic moduli and $\sigma_0$ represents the yield stress. The spatial assignment of random $E$ and/or $\sigma_0$ is a field of independent identically distributed (i.i.d.) random variables. That is, $\mathbf{\Theta} = \{E, \sigma_0\}$ is a strict-white-noise field, clearly a non-fractal. The mean values taken are those of aluminum: $E = 71\,\text{GPa}$, $\sigma_0 = 137\ \text{MPa}$, with the Poisson ratio $\upsilon = 0.348$ (Taylor, 1995).

**Model 2.** Anisotropic polycrystalline aggregates with random orientations. For individual crystals the elasticity tensor $\mathbf{D}^p$ and the yield tensor $\mathbf{\Pi}^p$ are given by

$$\begin{aligned} \mathbf{D}^p_{ijkl} &= \mathbf{R}^p_{im}\mathbf{R}^p_{jn}\mathbf{R}^p_{kr}\mathbf{R}^p_{ls}\mathbf{D}^{ref}_{mnrs}, \\ \mathbf{\Pi}^p_{ijkl} &= \mathbf{R}^p_{im}\mathbf{R}^p_{jn}\mathbf{R}^p_{kr}\mathbf{R}^p_{ls}\mathbf{\Pi}^{ref}_{mnrs}. \end{aligned} \tag{11}$$

where $\mathbf{D}^{ref}$ and $\mathbf{\Pi}^{ref}$ are the referential elasticity and yield tensor, $\mathbf{R}^p$ is the rotation tensor associated with a grain of type $p$. Also in this model, the random orientations form a strict-white-noise field. The material orientations are taken to be uniformly distributed on a circle; this is realized by an algorithm of Shoemake (1992). Values of the reference material parameters are given in Table 1.

**Table 1:** Material parameters in Model 2.

| Material | Elasticity[a] (GPa) | | | Plasticity[b] | | | | |
|---|---|---|---|---|---|---|---|---|
| | $c_{11}$ | $c_{12}$ | $c_{44}$ | $\sigma_0$(MPa) | $\sigma_{11}/\sigma_0$ | $\sigma_{22}/\sigma_0$ | $\sigma_{33}/\sigma_0$ | $\sigma_{12}/\sigma_0$ |
| Aluminum | 108 | 62.2 | 28.4 | 137 | 1.0 | 0.9958 | 0.9214 | 1.08585 |

[a] Material properties for cubic elastic symmetry (Hill, 1951)

[b] Material properties for the quadratic anisotropic yield criterion (Taylor, 1995)

A numerical study of both models, in plane strain, is carried out by ABAQUS. We take a sufficiently large domain that comprises $256\times 256$ squared-shaped grains. Each individual grain is homogeneous and isotropic, its $E$ being constant and $\sigma_0$ being a uniform random variable up to $\pm 2.5\%$ about the mean. Other kinds of randomness are studied in Section 5. We apply shear loading through one of the three types of uniform BC:

$$\begin{aligned} &\text{Kinematic: } \varepsilon^0_{11} = -\varepsilon^0_{22} = \varepsilon,\ \varepsilon^0_{12} = 0, \\ &\text{Mixed: } \varepsilon^0_{11} = \varepsilon, \sigma^0_{22} = -\sigma, \varepsilon^0_{12} = \sigma^0_{12} = 0 \\ &\text{Traction: } \sigma^0_{11} = -\sigma^0_{22} = \sigma,\ \sigma^0_{12} = 0, \end{aligned} \tag{12}$$

all consistent with (2)-(5).

The equivalent plastic strain contour plots under different BCs are shown in Fig. 1 for both models on domains $64\times 64$ grains; these smaller domains are chosen because graphics on larger domains become visually too fuzzy. We can find that the shear bands are at roughly $45^0$ to the direction of tensile loading under various BCs. This is understandable since we apply shear loading with equal amplitude in both directions, while the material field is inhomogeneous, so the shear bands are not at $45^0$ exactly. Regarding this inhomogeneity, the plastic grains tend to form in a geodesic fashion so as to avoid the stronger grains (Jeulin *et al*., 2008). Note that the shear band patterns under the three BCs are different. They differ by stress concentration factors and rank in the following order in terms of BCs: kinematic, mixed and traction.

Figures 2(a,b) show constitutive responses of volume-averaged stress and strain under three BCs for both models. The responses of single grain homogenous phases are also given for a reference. First, the curves under different BCs almost overlap, showing that the ($256\times256$) domain is very close to the Representative Volume Element (RVE), i.e. the responses are almost independent of the type of BC (Ostoja-Starzewski, 2005, 2008). The response under mixed-orthogonal loading is bounded from above and below by kinematic and traction loadings, respectively, although this is difficult to discern in Fig. 2(b). Of course, domains as large as possible are needed to assess fractal dimensions.

The results in Figs. 2(a,b) can also be described by hierarchies of bounds for elastic-hardening plastic composites (Jiang *et al*., 2001; Ostoja-Starzewski, 2008), such as

$$\begin{aligned} &\left\langle \mathbf{S}_1^{Tt} \right\rangle^{-1} \le \ldots \le \left\langle \mathbf{S}_{\delta'}^{Tt} \right\rangle^{-1} \le \left\langle \mathbf{S}_{\delta}^{Tt} \right\rangle^{-1} \le \ldots \le \left( \mathbf{S}_\infty^{T} \right)^{-1} \\ &\equiv \mathbf{C}_\infty^{T} \le \ldots \le \left\langle \mathbf{C}_{\delta}^{Td} \right\rangle \le \left\langle \mathbf{C}_{\delta'}^{Td} \right\rangle \le \ldots \le \left\langle \mathbf{C}_1^{Td} \right\rangle, \quad \text{for all} \quad 1 \le \sigma' < \sigma \le \infty. \end{aligned} \tag{13}$$

Here $\left\langle \mathbf{C}_{\delta}^{Td} \right\rangle$ and $\left\langle \mathbf{S}_{\delta}^{Tt} \right\rangle$ are the apparent tangent stiffness and compliance moduli, respectively. The superscript $d$ (or $t$) indicates the case of displacement (or traction) BC. Another type of hierarchy that applies is in terms of energies, see eqn (15) in (Ostoja-Starzewski & Castro, 2003).

Note that the curves of heterogeneous materials are always bounded from above by those of the corresponding homogeneous materials. However, the difference in case of Model 2 is larger - the reason for this is that, while in Model 1 we use a material whose parameters are arithmetic means of the microstructure, in Model 2 we have to use a material with all crystalline grains aligned in one direction ($\mathbf{R}^p = \mathbf{I}$).

# 4. Fractal patterns of plastic grains

Figures 3(a, b, c, d) show elastic-plastic transition patterns in Model 2 for increasing stress $\sigma$ in traction BC. The figures use a binary format in the sense that elastic grains are white, while the plastic ones are black. The plastic grains form plastic regions of various shapes and size, and we estimate their fractal dimension $D$ using a “box-counting method”

(Perrier *et al.*, 2006). Results of box counts for Figs. 3(a, b, c, d) are shown in Figs. 4(a, b, c, d), respectively, where we plot the *ln-ln* relationship between the box number *Nr* and the box size *r*, respectively. With the correlation coefficients very close to 1.0 for all four figures, we conclude that the elastic-plastic transition patterns are fractal. The same type of results, except for the fact that the spread of plastic grains is initially slower under the traction BC, are obtained for two other loadings in Model 2 as well as all loadings in Model 1.

Figures 5(a, b) show evolutions in time of the fractal dimension (*D*) under different BCs for both material models. We find that the curves depend somewhat on a particular BC: in both models the fractal dimension *D* grows slower under the traction BC than under the mixed BC, and then the kinematic BC, which corresponds to the characteristics of strain evolution curves we discussed in the previous section. However, note that they share a common trend regardless of the loading applied: *D* tends to 2.0 during the transition, showing that the plastic grains have a tendency to spread over the entire material domain.

Furthermore, the dependencies of *D* on the volume averaged plastic strain under different BCs are almost identical in case of both models, Fig. 6. This is very similar to materials' constitutive responses - say, the volume averaged stress *vs.* strain - which are independent of BCs for sufficiently large domains in Fig. 2. Thus, *D* turns out to be a useful parameter in quantifying the evolution of elastic-plastic transitions in heterogeneous materials at and above the RVE level.

## 5. Further discussion of Model 1

Here we examine Model 1 under several kinds of material parameter randomness and various model assumptions. First, the sensitivity of transition patterns to the material's model randomness is investigated through comparisons in two scenarios:

**Scenario A:** Scalar random field of the yield limit, with three types of randomness:

A1 - Yield limit is a uniform random variable up to $\pm 2.5\%$ about the mean.

A2 - Yield limit is a uniform random variable up to $\pm 0.5\%$ about the mean.

A3 - Deterministic case: no randomness in the yield limit.

**Scenario B:** Random field of the yield limit and/or elastic moduli,, $\mathbf{\Theta}_p = \{E_p, \sigma_p\}$, with four cases:

B1 = A1.

B2 – Modulus is a uniform random variable up to $\pm 2.5\%$ about the mean.

B3 - Yield limits and moduli are uniform random variables $\pm 2.5\%$ about their means.

B4 - Yield limits and moduli are uniform random variables $\pm 0.5\%$ about their means.

Results for A1-A3 and B1-B4 are shown in Figs. 7 and 8, respectively. From Figs. 7(a-b) one can conclude that different random variables in the model configuration lead to different transition patterns; overall, a lower randomness results in a narrower elastic-plastic transition. Next, in Fig. 8 we observe the randomness in yield limits to have a stronger effect than that in elastic moduli. When these both properties are randomly perturbed, the effect is even stronger – both, in the curves of the average stress as well as the fractal dimension vs. the average strain.

A test of the robustness of results of Model 1 involves a comparison of the original material with two other cases: (i) a hypothetical material with parameters of the aluminum increased by factor 2 [ $E = 142\ \text{GPa}$ , $\sigma_0 = 274\ \text{MPa}$ ]; a material with parameters of mild steel in [ $E = 206\ \text{GPa}$ , $\sigma_0 = 167\ \text{MPa}$ ] (Taylor, 1995). Figure 9(a) illustrates the evolutions of $D$ with respect to plastic strain for these materials. One can find that the curves of material 1) and 2) are almost identical and bounded from above by that of material 3), which is understandable, since the first two materials have the same yield strain while for the latter one it is less than the two.

In order to demonstrate the influence of yield strain more clearly, we scale the plastic strain by material's yield strain and plot the results again in Fig. 9(b). The three curves are now practically identical. Note that, after scaling of yield strain, the constitutive responses of all variants of Model 1 are also reduced to one smooth stress-strain curve, which can be fitted by, say, $\sigma_{12} = (2k/\pi)\tan^{-1}\left(d_{12}/b\right)$, where $k$ is the yield stress in shear, while $b > 0$ models a smooth curve; for $b \to 0$, the smooth curve tends towards the line of perfect-plasticity. This curve may be bounded by the linear elasticity/perfect plasticity with yield strain equal to 1.0.

# 6. Conclusions

We consider elastic-plastic transitions in random, linear elastic/perfectly-plastic media, where the yield limits and/or elastic moduli are taken as non-fractal, random fields (in fact, fields of i.i.d. random variables). In particular, two planar models are studied: a composite with isotropic grains and a polycrystal with anisotropic grains having orientation-dependent elasticity and Hill's yield criterion. By setting up three types of loadings consistent with the Hill-Mandel condition, the stress-strain responses and fractal dimensions of evolving plastic regions are obtained by computational mechanics. Referring to the three questions raised in the Introduction of this paper, we find that:

(a) The elastic-plastic transition occurs as a fractal, plane-filling process of plastic zones in both heterogeneous (fundamentally non-fractal) material models – one with random, fluctuations in yield limit and/or elastic moduli, and another with randomly oriented anisotropic grains. The fractal dimension of plastic zones increases monotonically as the macroscopically applied loading increases, with kinematic BC in a strongest growth of $D$, followed by the mixed-orthogonal BC, and then by the traction BC.

(b) Very similar fractal patterns and stress-strain curves are exhibited by both, the composite made of locally isotropic grains and the polycrystalline aggregate made of anisotropic grains. As the randomness in material properties decreases towards zero in the first model, the elastic-plastic transition tends from a smooth curved bend in the effective stress-strain curve towards a sharp kink and this is accompanied by an immediate plane-filling of plastic zones. Of course, the limiting case of no spatial randomness does not physically exist, i.e. a homogeneous material is but a hypothetical, idealized model. Also note that, in the model with anisotropic grains, no sharp kink can be recovered unless all the grains acquire an identical orientation.

(c) The fractal character of plastic zones is robust under changes of the model such as the change of strength in random perturbations in material properties or a change in the mean elastic moduli and yield limits. At this point we can only conjecture that the plane-filling character becomes space-filling in three-dimensional settings, with simulations of the latter appearing to be barely within the reach of present day computers.

While this research was set in the context of elastic-perfectly plastic grains, future studies will have to show, among others, how hardening affects the plane filling of plastic zones, and how fractal patterns change from the present (metal-type) models to soils.

## Acknowledgement

This work was made possible by the NCSA at University of Illinois and the NSF support under the grant CMMI-0833070.

## References

Feder, J., 2007. *Fractals (Physics of Solids and Liquids)*, Springer.

Hazanov, S., 1998. Hill condition and overall properties of composites, *Arch. Appl. Mech.* **68**, 385-394.

Hill, R., 1951. The elastic behavior of a crystalline aggregate, *Proc. Phys. Soc.* **A65**,

349-354.
Hill, R., 1963. Elastic properties of reinforced solids: some theoretical principles, *J. Mech. Phys. Solids* **11**, 357-372.
Jeulin, D. Li, W. & Ostoja-Starzewski, M., 2008. On the geodesic property of strain field patterns in elasto-plastic composites, *Proc. R. Soc. A* **464**, 1217-1227.
Jiang, M., Ostoja-Starzewski, M. & Jasiuk, I., 2001. Scale-dependent bounds on effective elastoplastic response of random composites, *J. Mech. Phys. Solids* **49**, 655-673.
Mandelbrot, B., 1982. *The Fractal Geometry of Nature*, W.H. Freeman & Co.
Ostoja-Starzewski, M., 1990. Micromechanics model of ice fields - II: Monte Carlo simulation, *Pure Appl. Geophys*. **133**:2, 229-249.
Ostoja-Starzewski, M., 2005. Scale effects in plasticity of random media: Status and challenges, *Intl. J. Plast*. **21,** 1119-1160.
Ostoja-Starzewski, M., 2008. *Microstructural Randomness and Scaling in Mechanics of Materials*, Chapman & Hall/CRC Press.
Ostoja-Starzewski, M. and Castro, J. 2003. Random formation, inelastic response and scale effects in paper, *Phil. Trans. R. Soc. A* **361**(1806), 965-986.
Perrier, E., Tarquis, A.M. & Dathe, A., 2006. A program for fractal and multi-fractal analysis of two-dimensional binary images: Computer algorithms versus mathematical theory, *Geoderma* **134**, 284-294.
Podio-Guidugli, P., 2000. A primer in elasticity, *J. Elast*. **58**, 1-104.
Poliakov, A.N.B., Herrmann, H.J., Podladchikov, Y.Y. & Roux, S., 1994. Fractal plastic shear bands, *Fractals* **2**, 567-581.
Poliakov, A.N.B. & Herrmann, H.J., 1994. Self-organized criticality of plastic shear bands in rocks, *Geophys. Res. Lett.* **21**(19), 2143-2146 .
Ranganathan, S.I. & Ostoja-Starzewski, M. 2008. Scale-dependent homogenization of inelastic random polycrystals, *ASME J. Appl. Mech*. **75**, 051008-1-9.
Sahimi, M. & Goddard, J.D. 1986. Elastic percolation models for cohesive mechanical failure in heterogeneous systems, *Phys. Rev. B* **33**, 7848.
Sahimi. M. 2003. *Heterogeneous Materials II*. Springer, New York.
Saouma, V.E. and Barton, C.C., 1994. Fractals, fractures, and size effects in concrete, *J. Eng. Mech. ASCE* **120**, 835-855.
Shaniavski, A.A. & Artamonov, M.A., 2004. Fractal dimensions for fatigue fracture surfaces performed on micro- and meso-scale levels, *Int. J. Fracture* **128**, 309-314.
Shoemake, K., 1992. Uniform random rotations, in: D. Kirk (Ed.), *Graphics Gems* **III.** Academic Press,
Sornette, D. 2004. *Critical Phenomena in Natural Sciences*, Springer, New York.
Taylor, L., Cao, J., Karafillis, A.P. & Boyce, M.C., 1995. Numerical simulations of sheet-metal forming, *J. Mater. Process. Tech.*,**50**, 168-179.
Zaiser, M. Bay, K. & Hahner, P., 1999. Fractal analysis of deformation-induced dislocation patterns, *Acta. Mater*, **47**, 2463-2476.

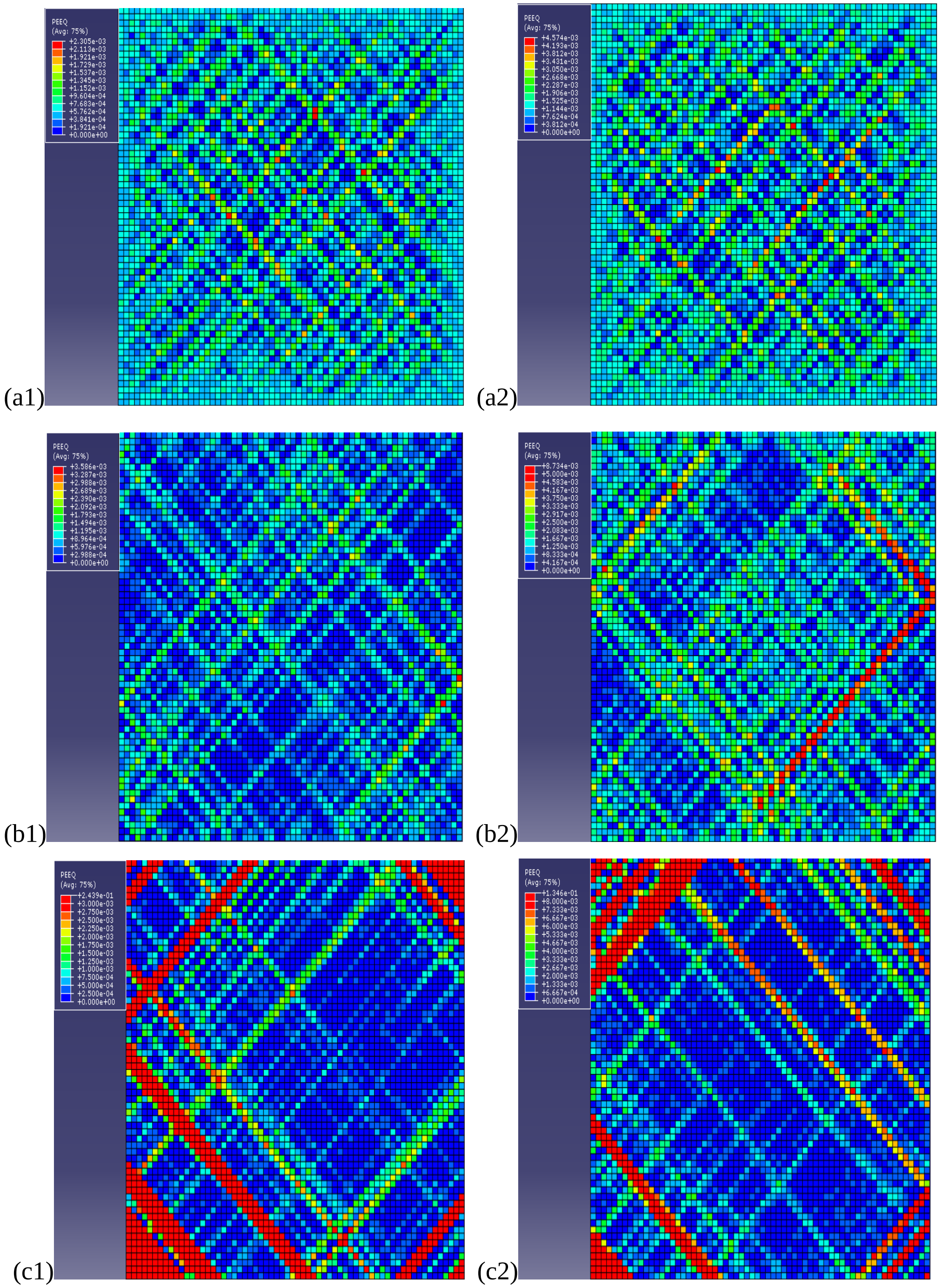

**Fig. 1** Plots of equivalent plastic strain on $64 \times 64$ domains for Model 1 (isotropic grains), and Model 2 (anisotropic grains) under various BCs: (a1,a2) kinematic (b1,b2) mixed (c1,c2) traction.

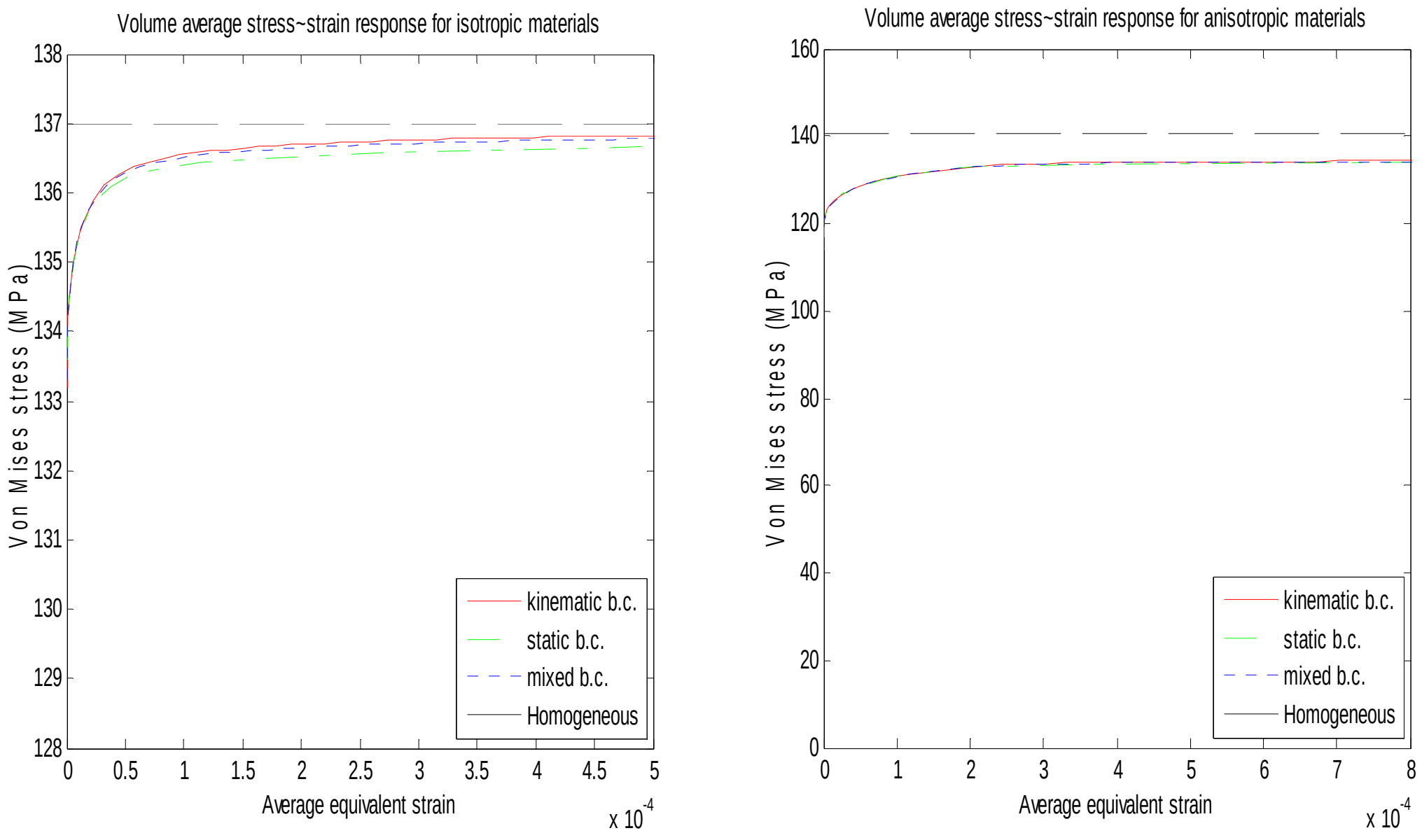

**Fig. 2** Volume-averaged stress~strain responses under different BCs for: (a) Model 1 (isotropic grains), and (b) Model 2 (anisotropic grains).

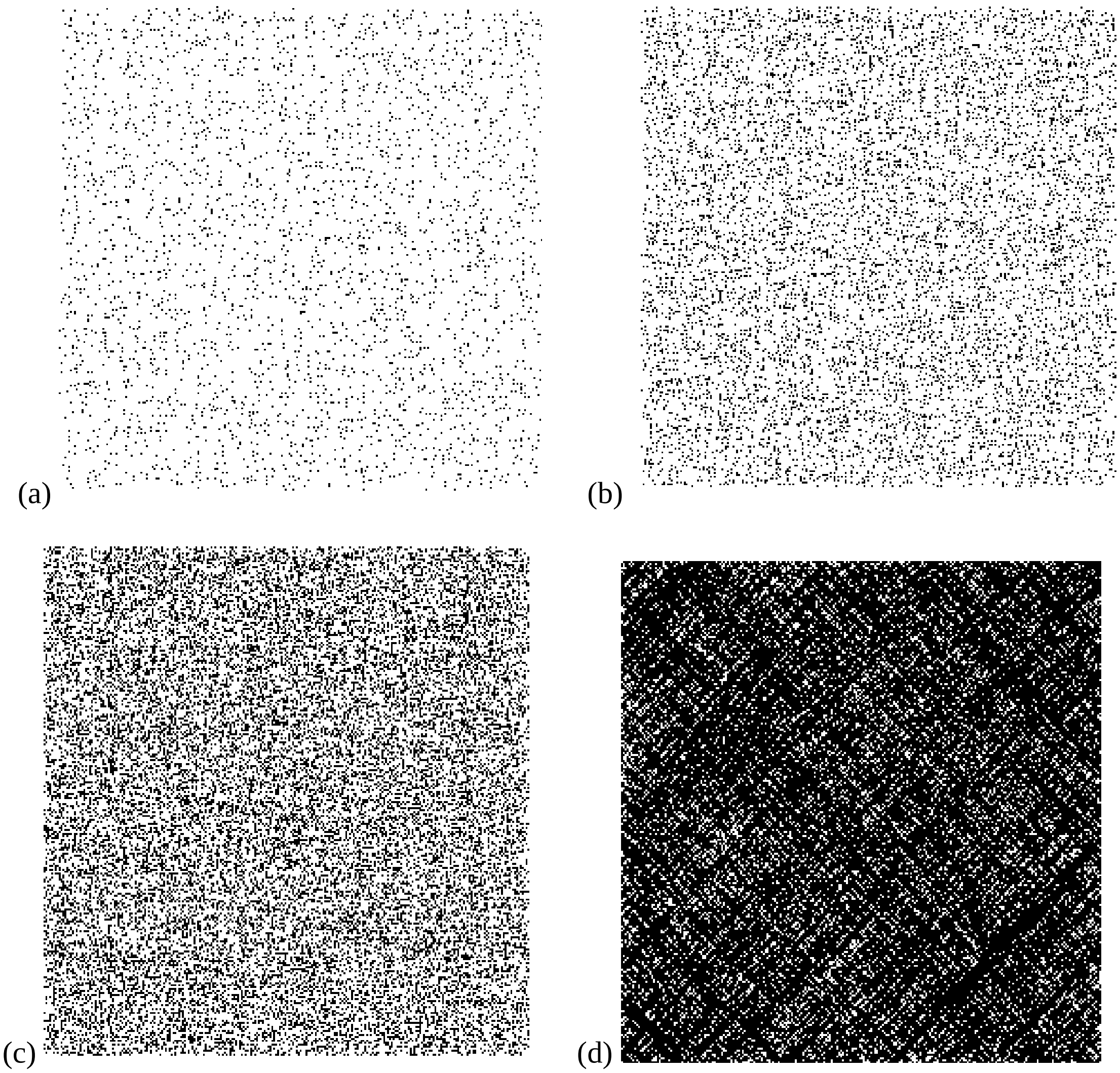

**Fig. 3** Field images (white: elastic, black: plastic) for Model 2 (anisotropic grains) at four consecutive stress levels applied via uniform traction BC. The set of black grains is an evolving set, with the fractal dimension given in Figs. 3(a, b, c, d),

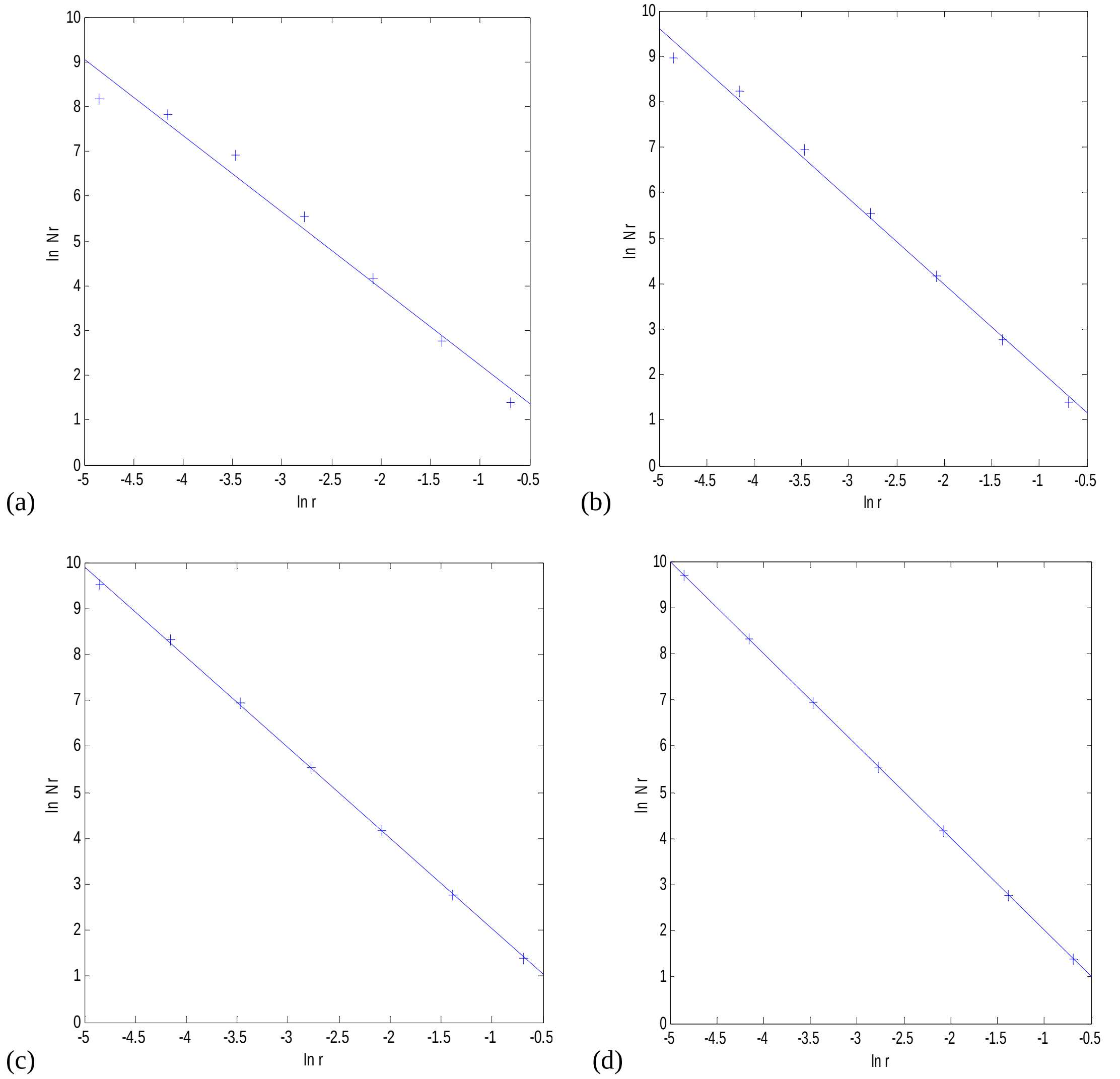

**Fig. 4** Estimation of the fractal dimension $D$ for Figs. 3(a, b, c, d), respectively, using the box-counting method: (a) $D = 1.667$, (b) $D = 1.901$, (c) $D = 1.975$, (d) $D = 1.999$. The lines corresponds to the best fit curve of $ln(Nr)$ vs. $ln(r)$.

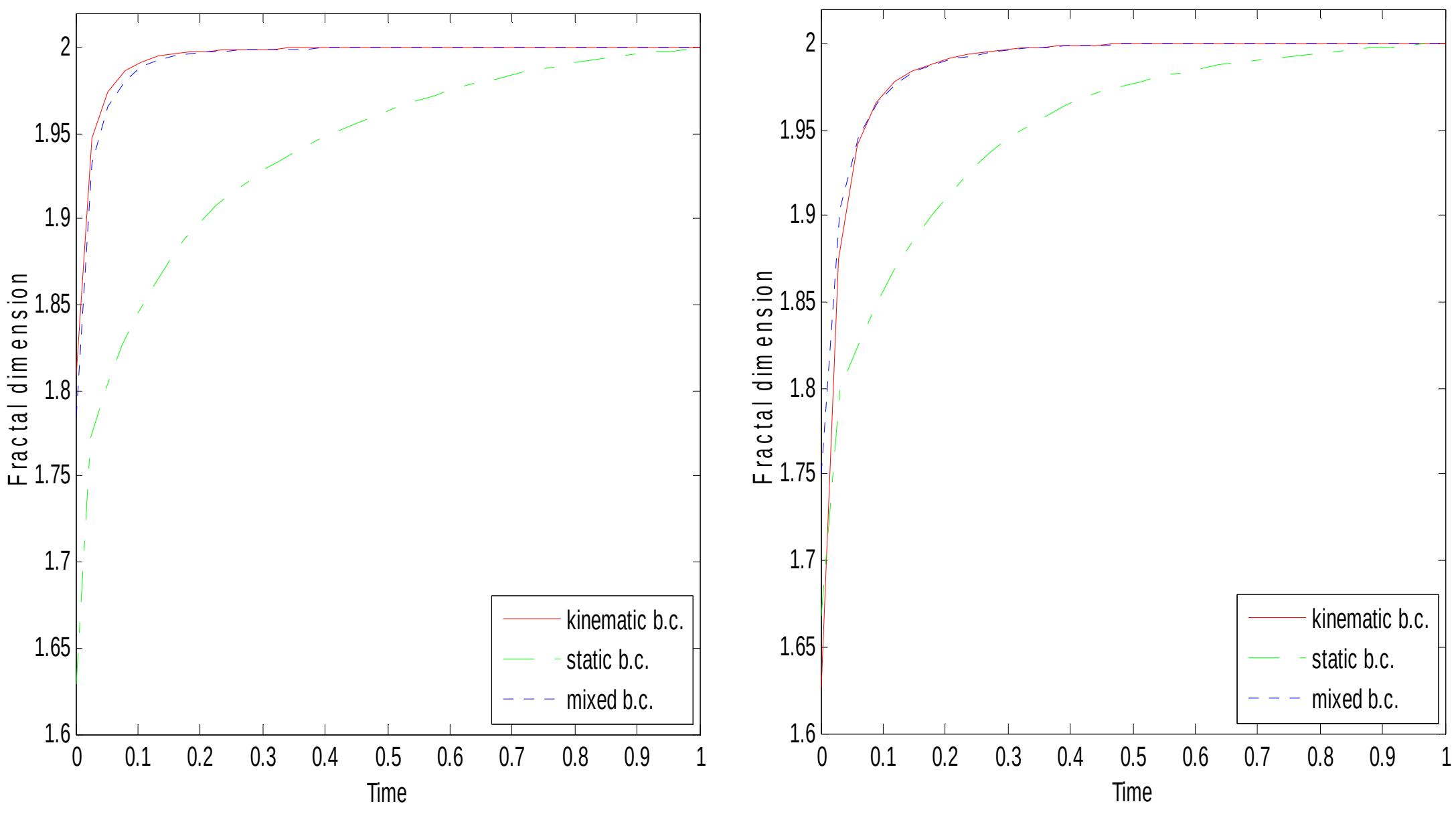

(a) (b)

**Fig. 5** Time evolution curves of the fractal dimension under different BCs for: (a) Model 1 (isotropic grains), and (b) Model 2 (anisotropic grains). All loadings are linear in time.

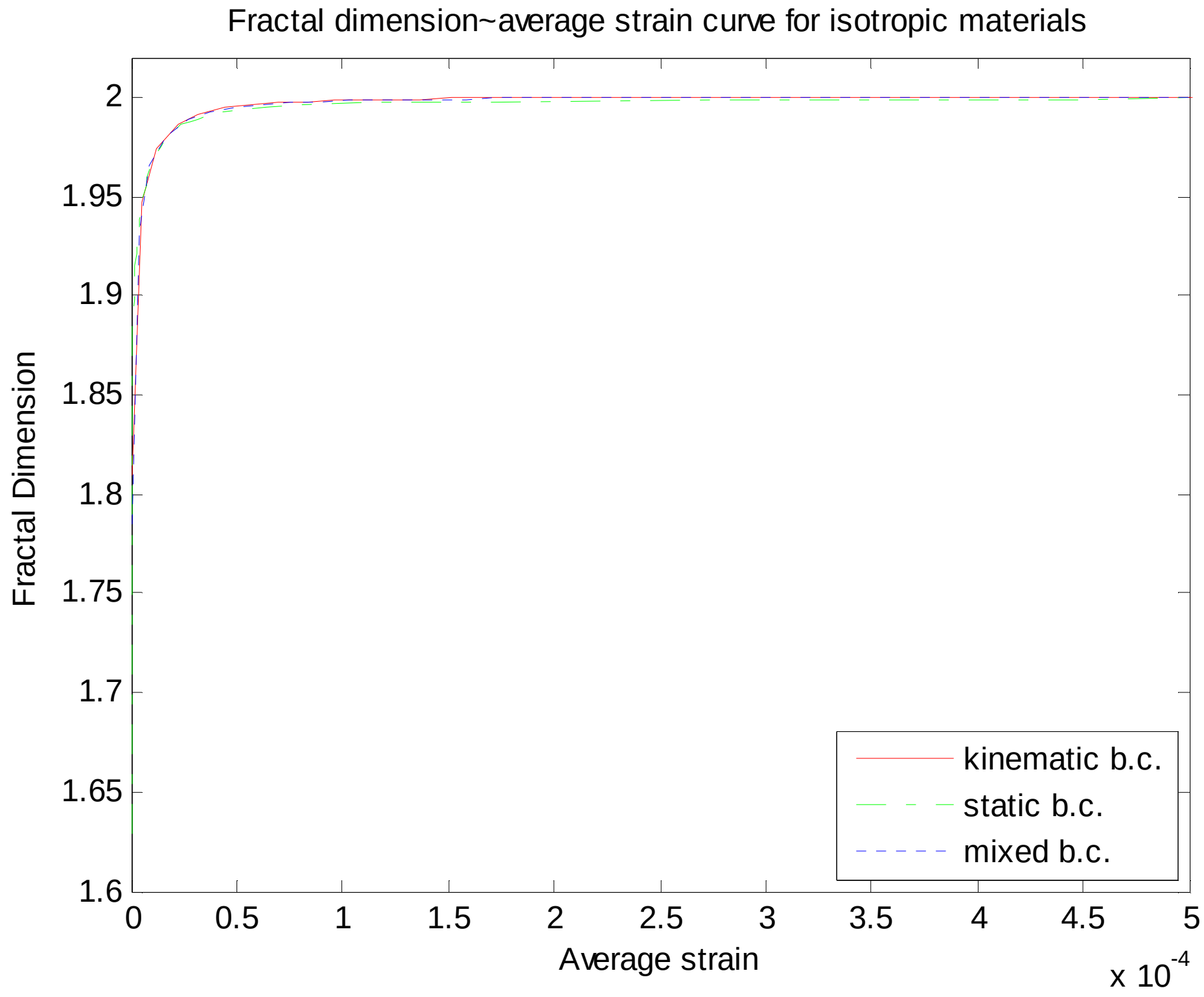

(a)

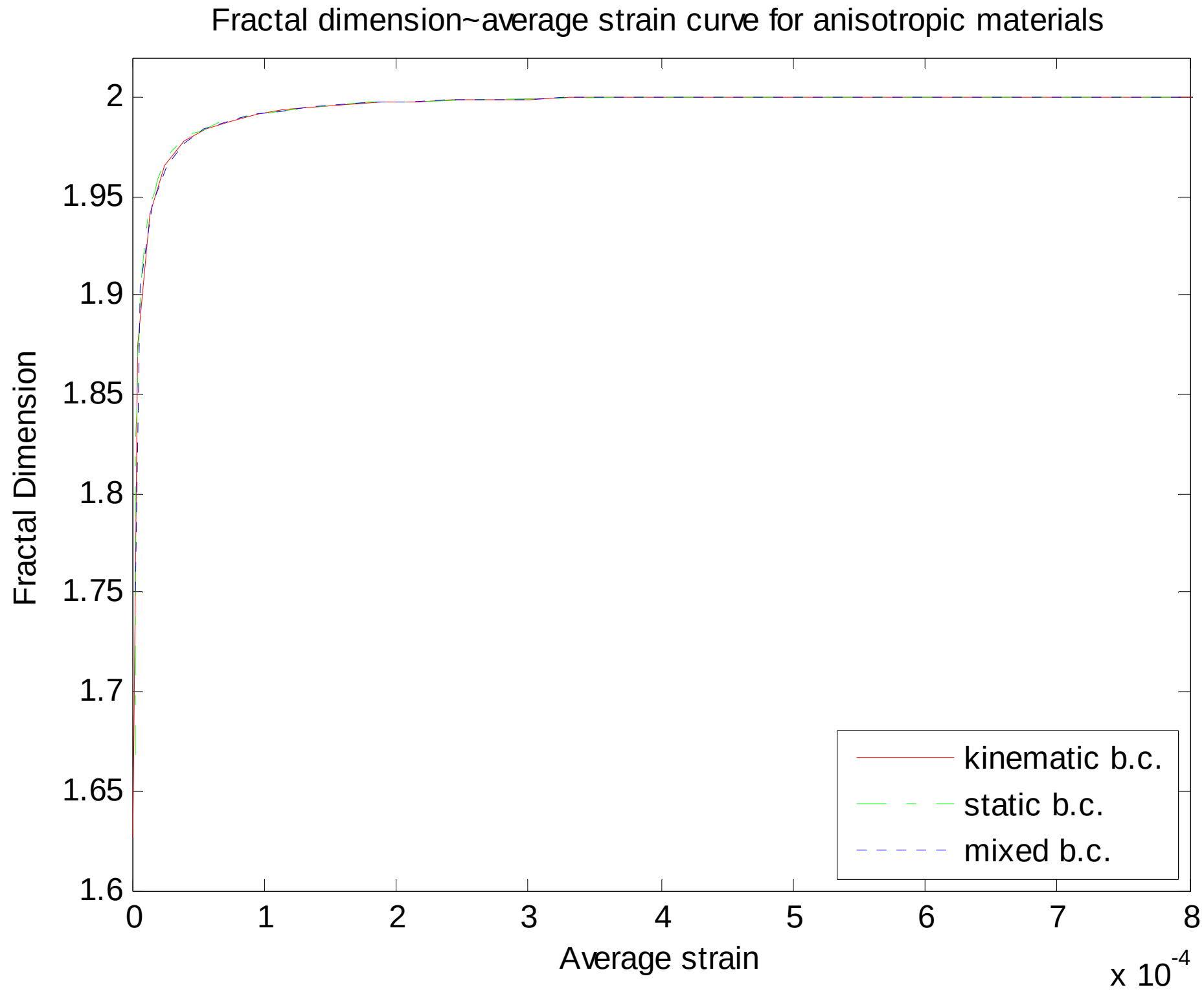

(b)

**Fig. 6** Fractal dimension~average strain curves under different BCs for: (a) Model 1 (isotropic grains), and (b) Model 2 (anisotropic grains).

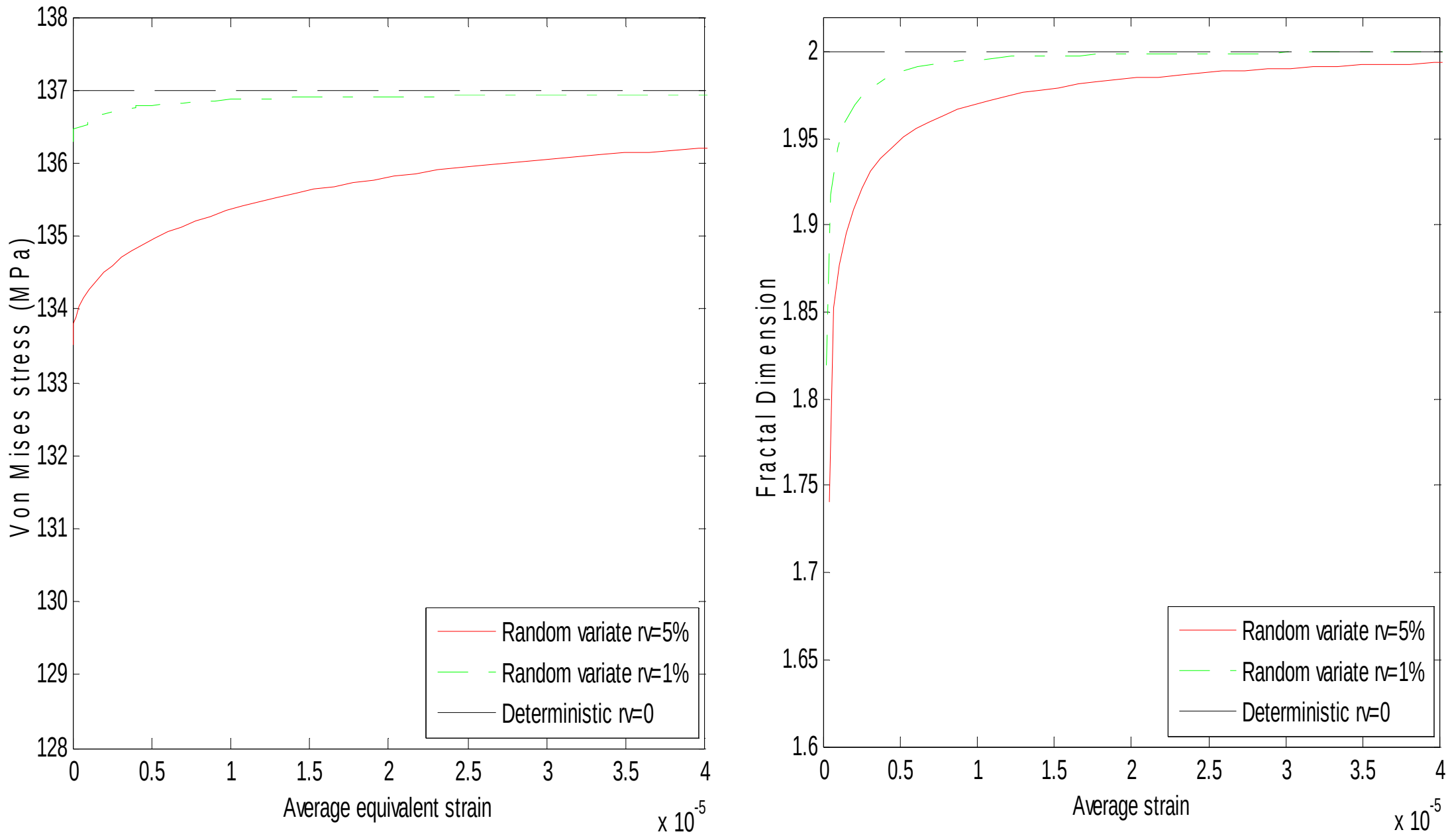

(a) (b)

**Fig. 7** Comparison by different random variants (RV=5%, 1% and 0-deterministic case): (a) Average stress-strain curves (b) Fractal dimension vs. average strain.

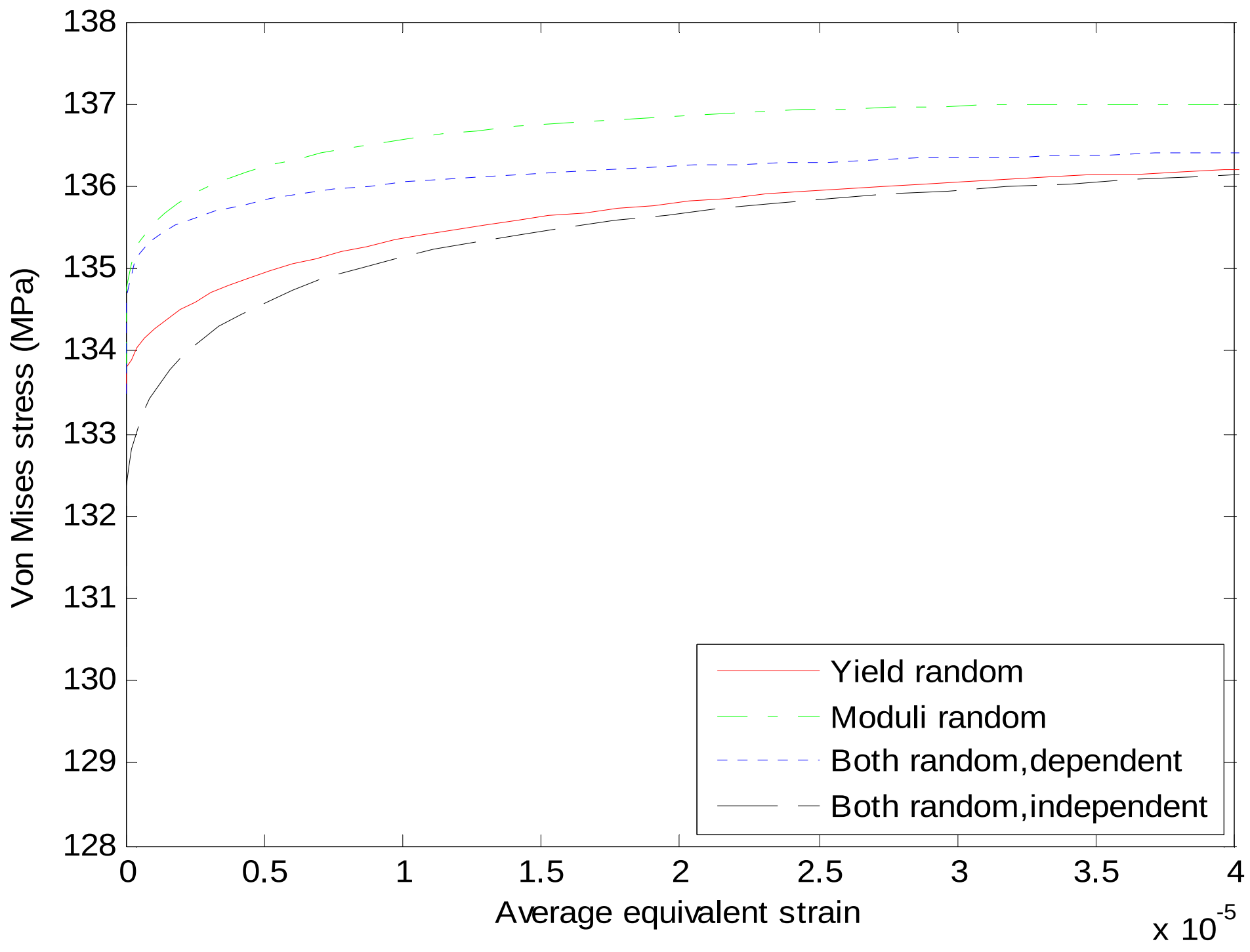

(a)

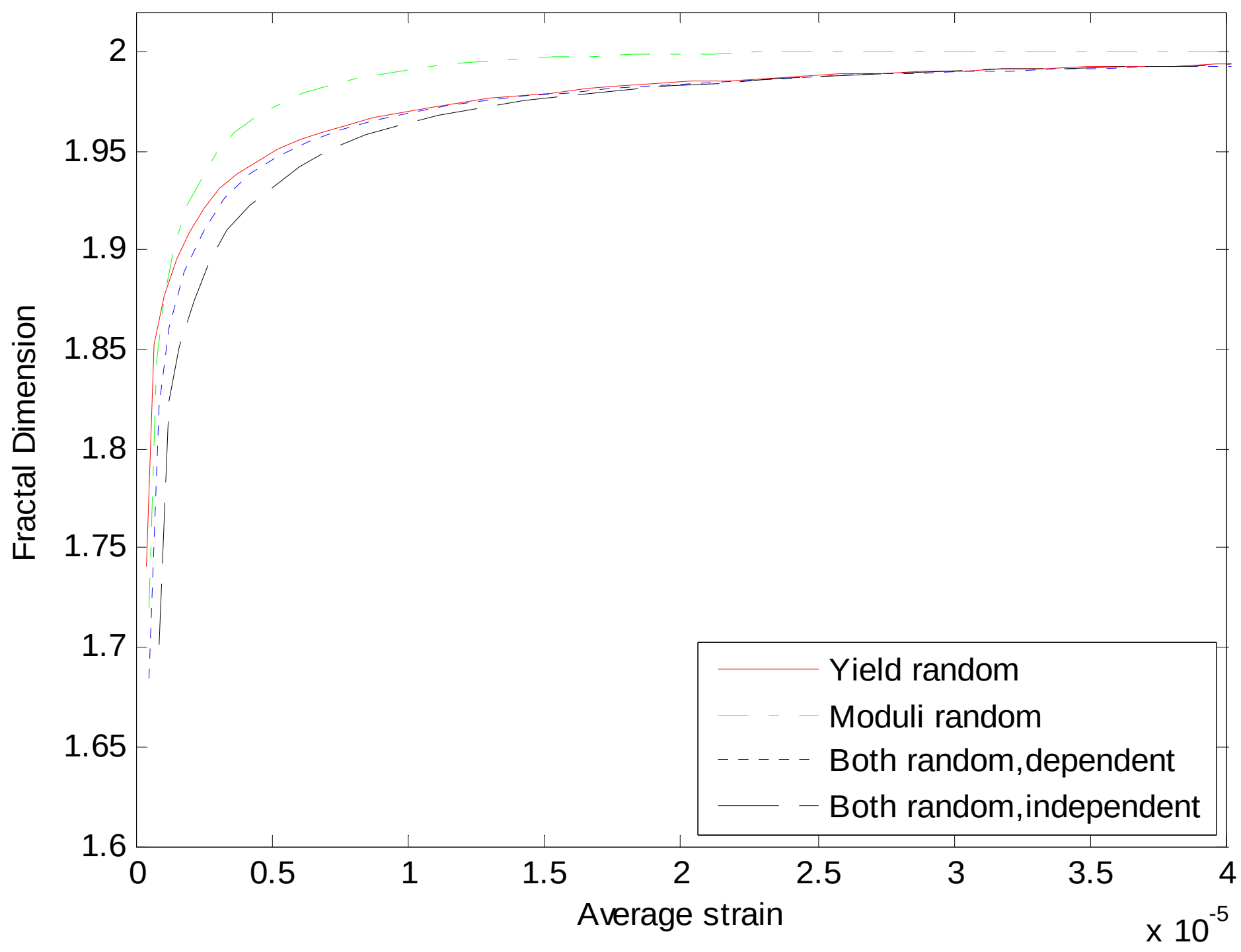

(b)

**Fig. 8** Comparison of the effects of random perturbations in the yield limit and/or elastic moduli: (a) average stress vs. average strain; (b) fractal dimension vs. average strain.

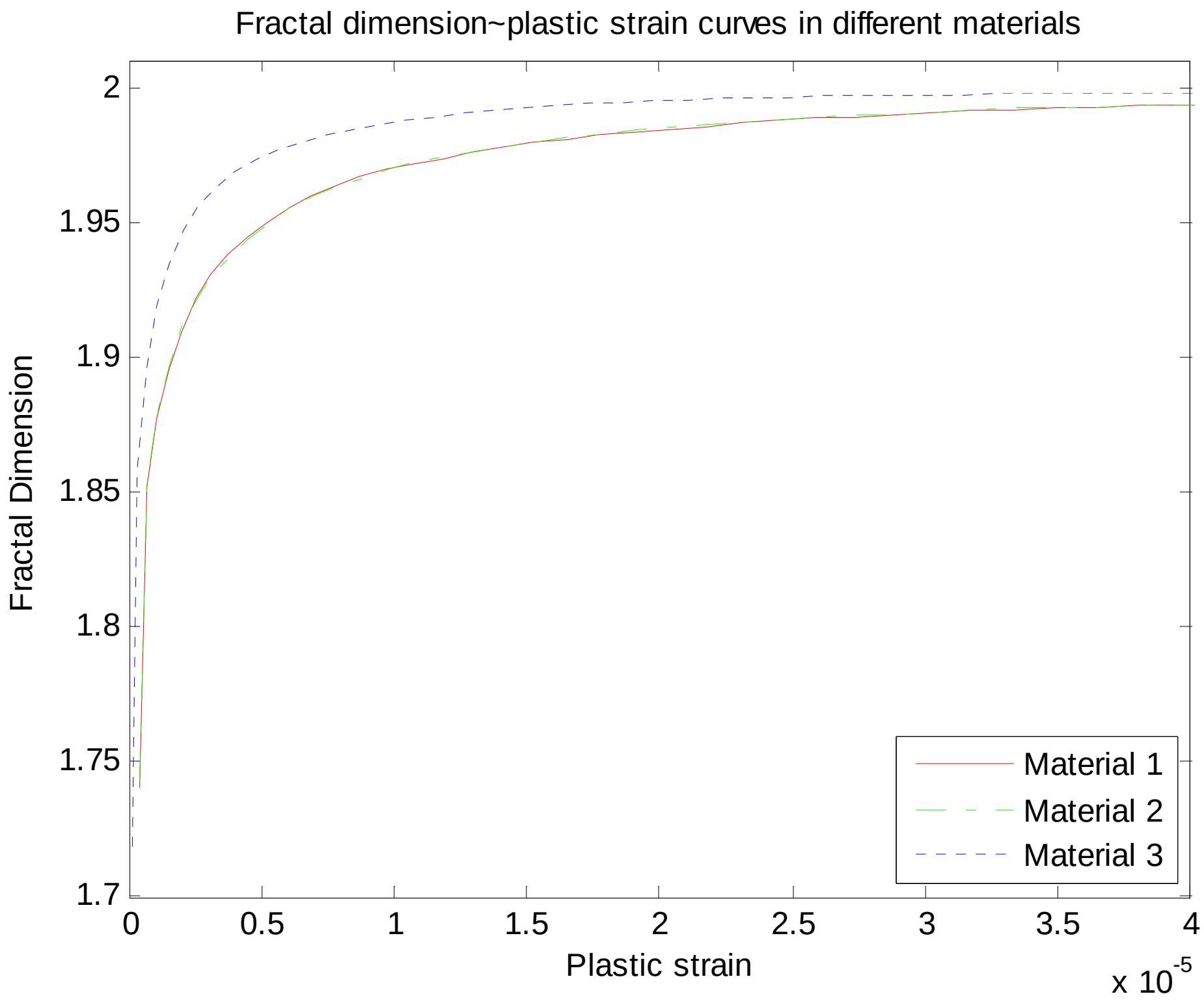

(a)

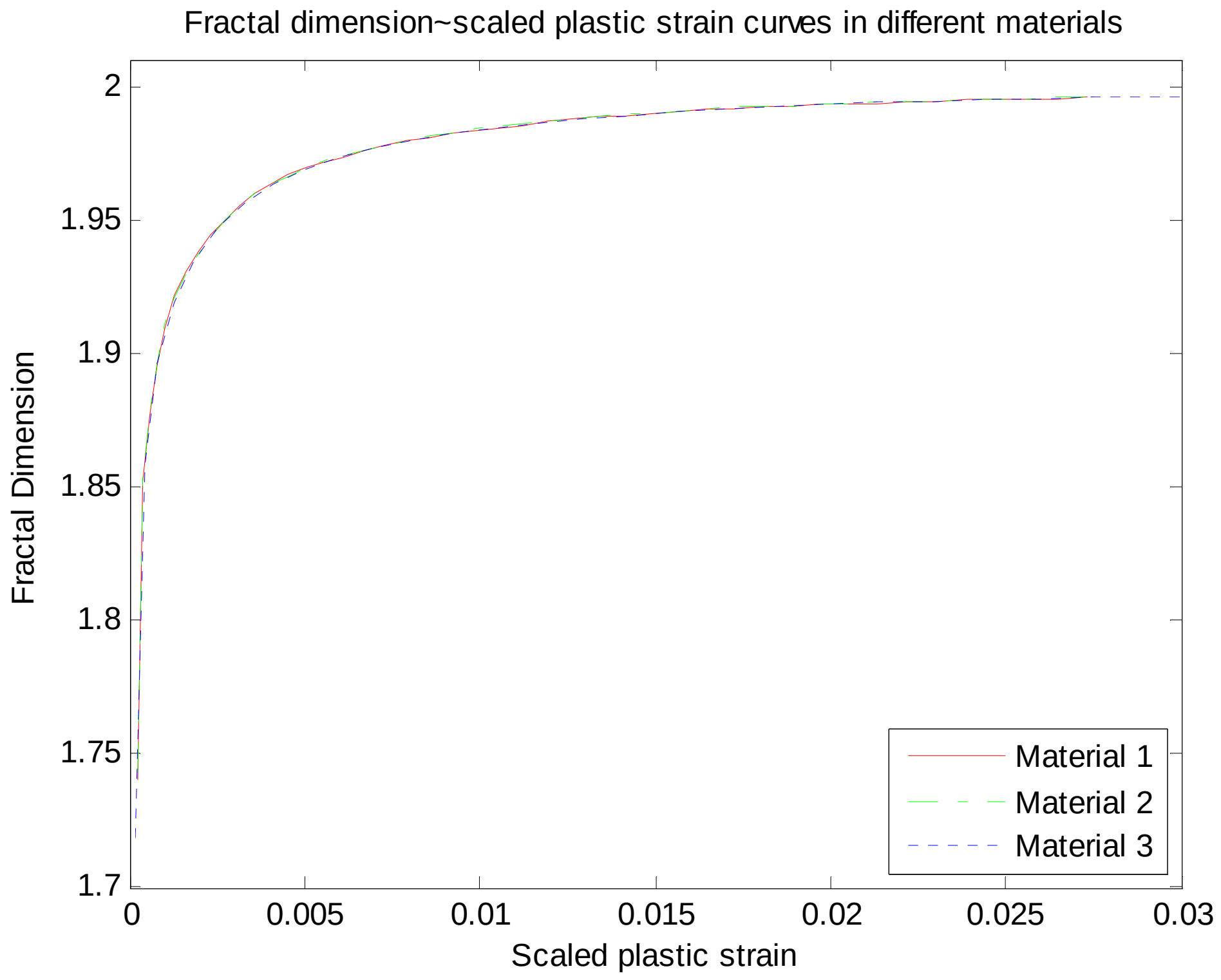

(b)

**Fig. 9** Comparison of different materials' responses: (a) Fractal dimension vs. plastic strain; (b) Fractal dimension vs. scaled plastic strain (i.e. scaled by yield strain).